# Atomic, electronic and magnetic structure of graphene/iron and nickel interfaces: theory and experiment

D.W. Boukhvalov,[1,2,*] Yu. N. Gornostyrev,[3,4] M. A. Uimin,[4] A. V. Korolev[4] and A. Ye Yermakov[4]

[1]*Department of Chemistry, Hanyang University, 17 Haengdang-dong, Seongdong-gu, Seoul 133-791, Korea*

[2]*Theoretical Physics and Applied Mathematics Department, Ural Federal University, Mira Street 19, 620002 Ekaterinburg, Russia*

[3]*Institute of Quantum Materials Science, Ekaterinburg 620175, Russia*

[4]*Institute of Metal Physics, Russian Academy of Sciences-Ural Division, 620041 Ekaterinburg, Russia*

*First-principles calculations of the effect of carbon coverage on the atomic, electronic and magnetic structure of nickel and iron substrates demonstrate insignificant changes in the interatomic distances and magnetic moments on the atoms of the metallic substrates. The coverage of the iron surface by mono- and few-layer graphene induces significant changes in the orbital occupancies and exchange interactions between the layers in contrast to the case of a nickel substrate for which changes in the orbital ordering and exchange interactions are much smaller. Experimental measurements demonstrate the presence of ferromagnetic fcc-iron in Fe@C nanoparticles and the superparamagnetic behavior of Ni@C nanoparticles.*

E-mail: danil@hanyang.ac.kr

# 1. Introduction

Graphene is a novel material proposed for its various electronic, spintronic, supercapacitative, catalytic and other applications [1-12] that require detailed surveys of various types of graphene based materials and its properties. The growth of graphene by chemical vapor deposition over transition metal substrates is proposed as the most promising method for the fabrication of large-scale and cheap graphene for further industrial application. [1,13-17] Graphene/transition metal interfaces are also suggested as possible perfect spin-filters [18] and magnetoresistive junctions. [19] The unusual catalytic properties of these systems have also been discussed experimentally and theoretically. [20-22] The employment of graphene films for the protection of metal surfaces from oxidation is also a focus of current research. [23-26] Exploring the interactions of graphene with metallic contacts is also an important subject for the development of graphene-based devices. [27-29]

Previous theoretical works have considered only the changes in the electronic structure of graphene over transition metal (further TM) substrates for spin filtering, doping of graphene from metals [30-33], and changes in the electronic structures and chemical activity of graphene [34-37] and carbon nanotubes [38] in the presence of metallic substrates [20-22] or adatoms. [32,33] The changes in the electronic and magnetic structure of a metal surface at a junction with a carbon layer have been beyond the scope of current experimental and theoretical studies. Recent experiments have reported changes of the magnetic properties of the metal core in TM-filled carbon nanotubes and carbon-coated TM-nanoparticles. [39-41] A detailed survey of the electronic structure of these compounds by soft X-ray measurements demonstrates that the electronic and crystal structures of the TM-cores is nearly the same as for the bulk compounds of these metals. [42] Significant changes in the magnetic properties of the TM/carbon interfaces without detected changes in the crystal and electronic structure of the metals require a systematic study of the structural, electronic and magnetic properties of the TM/carbon interfaces. Robust and persistent changes in the magnetic interactions in gamma-iron in the vicinity of interstitial carbon impurities [43] motivate us to examine theoretically and experimentally the probable changes in the magnetic properties of transition metal surfaces after carbon coating.

To test the influence of a carbon coating on the magnetic properties of TM surfaces, the (111) surface of fcc iron and nickel were chosen. This choice of the metallic substrates is due to practical reasons: the nickel surface is the first reported metal substrate for the growth of a single layer of graphene, [44,45] and the (111) surface of fcc iron is also reported as a substrate for the graphene [46]; additionally, the special role of the (111) surface of austenite (fcc-iron) in process of Widmanstätten ferrite formation has been discussed in the literature. [47,48] We also examined the cases of mixed-metal substrates (for example, iron single and double layers over nickel) to more deeply understand the impact of the surface and subsurface layers on the magnetic properties of carbon-coated metallic substrates. For the experimental survey, we used Fe@C and Ni@C nanoparticles, the composition of which is a metallic core covered by almost planar few-layer graphene with a small amount of defects (for the details of synthesis and characterization of these nanoparticles see Res. [42,49]).

## 2. Computational and experimental details

To imitate the graphene/TM interfaces, we employed a standard model of these systems [18, 20-22,30,31], a 2×2 supercell of graphene (8 carbon atoms in layer) above 9 layers of metals with four metal atoms in each layer (see inset of Fig. 1). The slabs are within periodic boundary conditions, continuous along x and y directions and separated from the next slab by 20 Å of empty space along z axis. The modeling was performed by density functional theory (DFT) in the pseudopotential code SIESTA, [50] as was done in our previous works. We used Siesta-3.2 version. [51] All calculations were performed using the local density approximation (LDA) [52] with spin polarization that has been employed previously to model graphene/metal interfaces. [18,30,31] Full optimization of the atomic positions was performed. During the optimization, the ion cores are described by norm-conserving nonrelativistic pseudopotentials [53] with cut off radii of 2.00, 2.05 and 1.25 a.u. for Fe, Ni and C, respectively, and the wavefunctions are expanded with a double-ζ plus polarization basis of localized orbitals for all atoms. Optimization of the force and total energy was performed within an accuracy of 0.04 eV/Å and 1 meV/atom, respectively. All calculations were performed with an energy mesh cut-off of 300 Ry and a k-point mesh of 4×4×2 in the Monkhorst-Pack scheme. [54] Because studied slabs contain few layer number of k-ponts in z direction is chosen more than one. A further increase in the energy mesh cut-off to 400 Ry and the k-point mesh to 18×8×4 provide changes in the total energy of the system less than 10 meV. The values of the exchange interactions between the metallic layers are defined as the difference between the total energies of the ferromagnetic and antiferromagnetic (when all spins in the layers above align upward and all spins in the layers below align downward) configurations of the spins ($E_{exch} = (E_{FM} - E_{AFM})/N$ where N is the number of metallic atoms in layer).

Mössbauer investigations of the Fe@C composites were performed using iron with a natural content of the $^{57}$Fe isotope. [55] For measurements of the Mössbauer spectra of the Ni@C nanocomposites, the nickel used for the preparation of the nanocomposite was preliminarily alloyed with 3 wt % $^{57}$Fe. Mössbauer spectra were recorded on an MS-2201 electrodynamic-type spectrometer in the absorption geometry using a resonant detector at a temperature of 300 K. The source of γ-radiation was the $^{57}$Co(Cr) isotope with an activity of 50 mCi. The mathematical processing of the experimental spectra was performed with the MS TOOL software [56]. The magnetization curves of the composites Fe@C and Ni@C were investigated in pulsed magnetic fields up to 35 T at a temperature of 77 K and on a SQUID magnetometer in magnetic fields up to 50 kOe at 300 and 2 K.

**3. Structural changes at the graphene/metal interface**

**3.1. Theory**

The first step of our survey is to check for changes in the crystal structure of the few-layer graphene and the TM surfaces produced by the formation of the interface. The results of the calculations demonstrate negligible changes in the graphene and nickel lattice parameters in the plane parallel to the junctions. For the check of the role of the choice of functional we performed the calculations within GGA with van der Waals corrections implemented in SIESTA code. [51, 57] Results of our calculations and comparison with recent theoretical results (Table I) demonstrate that the main difference between LDA and GGA+vdW approaches is in the values of binding energies. Binding energies calculated within LDA is almost two times bigger than obtained within GGA+vdW calculations. Despite this difference in the values of binding energies calculated by the two methods, obtained values are an order of magnitude larger than binding energies between layers in graphite. This suggests the formation of chemical bonds between metallic substrate and carbon cover. This significant difference in binding energies provides rather insignificant (about 3%) changes in graphene-substrate distances and negligible changes in the values of magnetic moments and orbital occupancies. Thus we can conclude that the choice of computational method (LDA or GGA+vdW) play no role for the exploration of magnetic properties of metal/graphene interface.

**Table I** Binding energies per carbon atom (in eV) and distances between graphene and metallic substrate (in Å) for the case of iron and nickel scaffolds calculated by different methods.

| System | LDA | | GGA+vdW | |
|---|---|---|---|---|
| | $E_{bind}$ | d | $E_{bind}$ | d |
| C/Fe | 0.16 | 2.07 | 0.10 | 2.14 |
| C/Ni | 0.17 | 2.05 | 0.12 | 2.11 |
| Ref. 58 | 0.18 | 2.04 | 0.11 | 2.12 |
| Ref. 59 | 0.20 | 2.00 | 0.12 | 2.10 |

The presence of the metallic substrate increases the interlayer distance in bi- and six-layer graphenes less than 7%. For the number of carbon layers, above five calculated interlayers, the distances remain nearly identical to the value for graphite. The changes in the interlayer distances of the metals slabs in the vicinity of graphene also vary within 2-3% near the values for the uncovered surfaces and are visible up to a 6% increase, only for the opposite side of the iron slab (Fig. 1a). However, these deviations from the case of a pure surface are caused by the limitation of the number of layers in our model and play no role for realistic metal surfaces that contain hundreds of layers. But observed light deviation of iron layers on opposite side play no valuable effects for electronic structure (Fig. 2j) and exchange interactions (Fig. 3a,d). Another important feature is the retention of the wiggling of the values of interlayer distances in uncovered metals after the formation of the graphene/TM interface. Similarly with the previous works [59] we also found decreasing of magnetic moments of topmost atoms of metallic substrates after coverage by graphene. Thus, from our calculations, we can conclude negligible changes in the atomic structure of the metal surfaces after coverage by mono- and few-layer graphene.

## 3.2. Experiment

In the case of Fe@C nanoparticles, the metallic core can consist of various phases. To check the atomic structure, we measured Mössbauer spectra of the Fe@C samples, which are shown in Fig. 4. The spectrum is a superposition of sextets, a doublet, and a paramagnetic line. After subtracting the sextet of $Fe_3C$ and the Fe–C doublet from the experimental spectrum, we calculated the hyperfine field distribution function P(H) shown in Fig. 4c from the difference spectrum (Fig. 4b). The negative isomer shift of the paramagnetic component IS = −0.09 mm/s, as noted in the literature, [60-62] is characteristic for γ-Fe(C) rather than for the superparamagnetic particles α-Fe. In addition to this paramagnetic component the peak with $H_{hf}$ = 330 kOe, which corresponds to α-Fe particles in the ferromagnetic state, there is a broad distribution of hyperfine fields, which can correspond both to solid solutions of carbon in iron Fe(C) with various carbon concentrations and to the iron carbides $Fe_xC_y$. The sextet with Hhf = 208 kOe probably corresponds to the iron carbide $Fe_3C$. We also cannot exclude the formation of a small amount of other carbides $Fe_xC_y$. The phase state of the nanocomposite based on Fe@C is quite complex and can be described by different sets of phases, taking into account the formation of carbides, including metastable phases, and the formation of α-iron and high-temperature γ-iron phase, which is capable of forming metastable solid solutions with a high carbon concentration in the nanostate.

## 4. Electronic and magnetic structure of the graphene/metal interfaces

### 4.1. Experiment

To verify the results of the calculations, we performed measurements of magnetic properties of the Fe@C and Ni@C nanoparticles. The results of the measurements demonstrate the pure ferromagnetic behavior of carbon-encapsulated iron nanoparticles without any traces of antiferromagnetic phases (Fig. 5) and superparamagnetic behavior for the case of the nickel core (Fig. 6). To check for the possible existence of an antiferromagnetic phase, the shift of the hysteresis loop was monitored when the samples were cooled in a magnetic field. Coexistence of ferromagnetic and antiferromagnetic or ferromagnetic and ferrimagnetic phases provide the shift of hysteresis loops. Measurements of the magnetization curves performed for the cooling from 400 to 5 K first in a field of +5 T and then in a field of -5 T in the limited cycle, and later at -1T and -0.5 T in a partial cycle, does not induce any shifts in the hysteresis loop (Fig. 5). Similarly, the measurements of the temperature dependence M(T) did not identify specific contributions from secondary magnetic phases (Fig. 6a). In the case of the presence of the antiferromagnetic secondary phase after obtaining the Neel temperature, the magnetization should drop. However, we do not observe any singularities for the temperatures from 2 to 300 K. In the case of the Ni@C nanoparticles, we observe superparamagnetic behavior in agreement with the Langevin formula at room temperature and ferromagnetic behavior for temperatures below 77 K.

## 4.2. Theory

As discussed in the previous paragraph, insignificant changes in the interlayer distances also provide insignificant changes in the values of the magnetic moments of the metal substrates (Fig. 1b) and lead to visible changes in the density of states only in the two iron layers nearest to the graphene (Fig. 2b,c). Surprisingly, these insignificant changes in the lattice parameters and the densities of states of the iron substrate before and after coverage by graphene coexist with dramatic changes in the exchange interactions in the iron substrate (Fig. 3a). In contrast with iron, similar insignificant changes in the atomic structure of the nickel substrate after coverage by carbon layers does not provide valuable changes in the magnetic properties (Fig. 3d).

The influence of orbital ordering for the exchange interactions was deeply explored for metal-oxygen based systems. [63,64] To explore this effect, we check the Mulliken orbital occupancies and find correlations between the changes in the orbital occupancies and variations in the exchange interactions. Both iron and nickel is in high spin states. There are five electrons with parallel spins (spin-up) and one for iron and three for nickel with antiparallel (spin-down). The number of layer and presence of carbon cover lead negligible effects for the 'spin-up' electrons and valuable changes for 'spin-down' electrons. Further we will discuss only 'spin-down' electrons of 3d orbitals of iron and nickel. In the case of the iron surface covered by carbon layers, a significant increasing of the occupancies of the $z^2$ orbitals in the top layers (this change provides visible changes in the density of states picture, see Fig. 2), decay of occupancies of $x^2$-$y^2$ oirbitals in all layers and an increase in the occupancy of the yz orbitals in all layers were observed (see Fig. 7). The magnitudes of the oscillations of the occupancies of the two other orbitals (xz and xy) are negligible and omitted from the discussion for clarity. Additionally, we can see that the coverage by mono- and bilayer graphene provides larger changes in the occupancies of the 3d orbitals in the iron substrate than for six-layer graphene. This effect is caused by the stronger coupling of mono- and bilayer graphene with the iron substrate and the partial decoupling of graphene with six or more layers (see Fig. 1a) for which the interlayer interaction is much stronger. [36] In the case of the nickel substrate, both the orbital occupancies and the interlayer distances remain nearly the same. The substitution of one or two top layers of iron by nickel retain the significant dependence of the magnetic properties of surface on the presence and number of layers of the carbon coating, in contrast to the case of substitution of one or two of the top layers of nickel by iron when the exchange interactions and the dependence from the carbon coating are almost the same as in the case of the pure nickel substrate. Thus, we note that the bulky part of the substrate determines the magnetic properties of graphene/metal interface and that one or two layers of other metals on these interfaces play a minor role.

The nature of this distinct difference between the iron and nickel substrates is the number of the electrons in *3d* shell. Our calculations report that all of the studied metals in the vicinity of the uncovered surface are in the high-spin configuration. In this configuration, iron has six electrons, five of which have parallel spins (spin-up), and the sixth of which is antiparallel (spin-down). The charge redistribution of fully occupied spin-up orbitals does not visible change after appearance of carbon coverage and further we will discuss only orbital distribution of spin-down electrons. In the absence of a graphene coating, these electrons on the atoms of the boundary layers of the slab are located on the $x^2$-$y^2$ orbital. Coverage of iron surface by graphene provides a change in the orbital ordering. Now, the sixth electron is distributed between the $z^2$ and yz orbitals, and the occupancy of the $x^2$-$y^2$ orbitals at all layers of iron decreases drastically for the case of mono- and bilayer graphene over the slab. This reordering of the orbital occupancies also occurs on the other side of the slab and provides changes in the exchange interactions between the layers (Fig. 3a) and increases the interlayer distance (Fig. 1a). Only in the case of a weaker bond with the metal substrate does the six-layer graphene occupancy of the $x^2$-$y^2$ orbitals of iron return to the patterns for the pure iron slab. Nickel has eight electrons in the 3d shell that provide five electrons with spin up and three electrons with spin down in the high-spin state and almost full occupancy of the $2z^2$, $x^2$-$y^2$ and yz orbitals. After coverage of the nickel substrate by graphene, we observe (Fig. 7) near absence of electron redistribution because all of the orbitals employed in the metal-graphene hybridization are already occupied. Hybridization between atoms of metals and between metal and carbon atoms explains why the bulky part of the substrates determines the magnetic behavior of the top few layers of the other metal with and without a carbon coating.

5. Conclusions

Based on experimental and theoretical results, we conclude that coverage of fcc iron by graphene conserved the fcc crystal structure and dramatically changed the magnetic properties of the iron substrate, whereas the magnetic properties remain nearly unchanged in the case of the nickel substrate. Increasing the number of graphene layers decreased the change in the magnetic properties between the coated and uncoated iron surfaces. The nature of these effects is related to the changes in the orbital ordering (especially, the swap of electrons between the $2z^2$ and $x^2-y^2$ orbitals) due to the formation of iron-carbon bonds and the absence of any changes in the orbital ordering of the nickel substrate.

**References**

1. A. K. Geim, *Science* 2009, **324**, 1530.
2. R. J. Young, I. A. Kinloch, L. Gong, and K. S. Novoselov, *Compos. Sci. Tech.* 2012, **72**, 1459.
3. W. Choi, I. Lahiri, R. Seelaboyina, and Y. S. Kim *Crit. Rev. Solid State Mater.* 2011, **35**.
4. H. Wang, T. Maiyalagan, and X. Wang, *ACS Catal* 2012, **2**, 781.
5. S. Dutta and S. K. Pati, *J. Mater. Chem.* 2010, **20**, 8207.
6. Y. Shao, J. Wang, H. Wu, J. Liu, I. A. Aksay, and Y. Lin, *Electroanalysis* 2010, **22**, 1027 ().
7. S. R. Vivekhand, C. S. Rout, K. S. Subrahmanyam, A. Govindraj, and C. N. R. Rao, *J. Chem. Sci.* 2008, **120**, 9.
8. C. L. Su, M. Acik, K. Takai, J. Lu, S.-J. Hao, Y. Zheng, P. P. Wu, Q. Bao, T. Enoki, Y. J. Chabal, and K. P. Loh, *Nat. Commun.* 2012, **3**, 1298.
9. M. J. Allen, V. C. Tung, and R. B. Kaner, *Chem. Rev.* 2010, **110**, 132.
10. S. Navalon, A. Dhakshinamoorrthy, M. Alvaro, and H. Garsia, *Chem. Rev.* 2014, **114**, 6179.
11. X. K. Kong, C. L. Chen, and Q. W. Chen, *Chem. Soc. Rev.* 2014, **43**, 2841.
12. Q. Tang, Z. Zheng, and Z. F. Chen, *Nanoscale* 2013, **5**, 4541.
13. C. Mattevi, H. Kim and M. Chhowalla, *J. Mater. Chem.* 2010, **21**, 3324.
14. D. Wei, B. Wu, Y. Guo, and Y. Liu, *Acc. Chem. Res.* 2013, **46**, 106.
15. Y. Zhang, L. Zhang, and C. Zhou, *Acc. Chem. Res.* 2013, **46**, 2329.
16. K. Yan, L. Fu. H. Peng, and Z. Liu, *Acc. Chem. Res.* 2013, **46**, 2263.
17. X. Li, et. al., *Science* 2009, **324**, 1312.
18. V. M. Karpan, P. A. Khomyakov, A. A. Starikov, G. Giovannetti, M. Zwierycki, M. Talanana, G. Brocks, J. van den Brink, and P. J. Kelly, Phys. Rev. B 2008, **78**, 195419.
19. O. V. Yazyev and A. Pasquarello, *Phys. Rev. B* 2009, **80**, 035408.
20. A. Ye. Yermakov, D. W. Boukhvalov, M. A. Uimin, E. S. Lokteva, A. V. Erokhin, and N. N. Schegoleva, *ChemPhysChem* 2013, **14**, 381.

21. V. Erokhin, E. S. Lokteva, A. Ye. Yermakov, D. W. Boukhvalov, K. I. Maslakov, E. V. Golubina and M. A. Uimin, *Carbon* 2014, **74**, 291.
22. D. W. Boukhvalov, Y.-W. Son, and R. S. Ruoff, *ACS Catal.* 2014, **6**, 2016.
23. S. Chen, et. al., *ACS Nano* 2012, **5**, 1321.
24. N. T. Kirkland, T. Schiller, N. Medhekar, and N. Birbilis, *Corros. Sci.* 2012, **56**, 1.
25. F. Zhou, Z. Li, G. J. Shenoy, L. Li and H. Liu, *ACS Nano* 2013, **7**, 6939.
26. M. Schriver, W. Regan, W. J. Gannett, A. M. Zaniewski, M. F. Crommie, and A. Zettl, *ACS Nano* 2013, **7**, 5763.
27. B. Huard, N. Stander, J. A. Sulpizio, and D. Goldhaber-Gordon, *Phys. Rev. B* 2008, **78**, 121402.
28. Varykhalov, M. R. Scholz, T. K. Kim, and O. Raeder, *Phys. Rev. B* 2010, **82**, 121101.
29. P. Blake, R. Yang, S. V. Morozov, F. Scheidin, L. A. Ponomarenko, A. A. Zhukov, R. R. Nair, I. V. Grigorieva, K. S. Novoselov, and A. K. Geim, *Solid Statae Commun.* 2009, **149**, 1068.
30. G. Giovannetti, P. A. Khomyakov, G. Brocks, V. M. Karpan, J. van den Brink and P. J. Kelly, *Phys. Rev. Lett.* 2008, **101**, 026803.
31. M. Bokdam, P. A. Khomyakov, G. Brocks, Z. C. Zhong, and P. J. Kelly, Nano Lett. 2011, **11**, 4631.
32. Q. Wang, D. X. Ye, Y. Kawazoe, and P. Jena, *Phys. Rev. B* 2012, **85**, 085404.
33. M. X. Liu et. al., *Nano Lett.* 2013, **13**, 137.
34. L. E. M. Steinkasserer, B. Paulus, and E. Voloshina, *Chem. Phys. Lett.* 2014, 148, 597.
35. E. Voloshina, R. Ovcharenko, A. Shulakov, and Y. Dedkov, *J. Chem. Phys.* 2013, **138**, 154706.
36. M. A. Kuroda, J. Tresoff, R. A. Nistor, and G. J. Martyna, *Phys. Rev. Appl.* 2014, **1**, 014005.
37. G. M. Sipahi, I. Žutić, N. Atodiresei, R. K. Kawakami, and P. Lazić, *J. Phys.: Condens. Matter* 2014, **26**, 104204.
38. W. Zhu and E. Kaxiras, *Nano Lett.* 2006, **6**, 1415.
39. J. Bao, Q. Zhou, J. Hong, and Z. Xu, *Appl. Phys. Lett.* 2002, **81**, 4592.
40. J. Bao, C. Tie, Z. Suo, Q. Zhou, and J. Hong, *Adv. Mater.* 2002, **14**, 1483.
41. A. Leonhardt, M. Ritschel, R. Kozhuharova, A.Graff, T. Mühl, R. Huhle, I. Mönch, D. Elefant, and C. M. Schneider, *Diam. Related Mater.* 2003, **12**, 790.
42. V. R. Galakhov, A. Buling, M. Neumann, N. Ovechkina, A. Shkvarin, A. Semenova, M. Uimin, A. E. Yermakov, E. Z. Kurmaev, O. Vilkov, and D. W. Boukhvalov, *J. Phys. Chem. C* 2011, **115**, 24615.
43. D. W. Boukhvalov, Yu. N. Gornostyrev, M. I. Katsnelson, and A. I. Lichtenstein, *Phys. Rev. Lett.* 2007, **99**, 247205.
44. K. S. Kim, et. al., *Nature* 2008, **457**, 706.
45. S. J. Chae, et. al., *Adv. Mater.* 2009, **21**, 2328.
46. Y. Xue, B. Wu, Y. Guo, L. Huang, L. Jiang, J. Chen, D. Geng, Y. Liu, W. Hu, and G. Yu, *Nano Res.* 2011, **4**, 1208.
47. J. D. Watson and P. G. McDougall, *Acta Metall.* 1973, **21**, 961.
48. A. Ali and H. K. D. H. Bhadeshia, *Mat. Sci. Tech.* 1990, **6**, 781.
49. V. R. Galakhov, S. N. Shamin, E. M. Mironova, M. Uimin, A. E. Yermakov, and D.W. Boukhvalov *JETP Lett.* 2013, **96**, 710.
50. J. M. Soler, E. Artacho, J. D. Gale, A. Garsia, J. Junquera, P. Orejon, and D. Sanchez-Portal, *J. Phys.: Condens. Matter* 2002, **14**, 2745.
51. C. F. Sans-Navarro, R. Grima, A. Garsia, E. A. Bea, A. Soba, J. M. Cela, and P. Odejon, *Theor. Chem. Acc.* 2011, **128**, 825.
52. J. P. Perdew and A. Zunger, *Phys. Rev. B* 1981, **23**, 5048.
53. N. Troullier and J. L. Martins, *Phys. Rev. B* 1991, **43**, 1993.
54. H. J. Monkhorstand and J. D. Pack, *Phys. Rev. B* 1976, **13**, 5188.
55. V. A. Tsurin, A. Ye. Yermakov, M. A. Uimin, A. A. Mysik, A. N. Schegoleva, V. S. Gaviko, and V. V. Maikov, *Phys. Solid State* 2014, **56**, 287.

56. V. S. Rusakov, Mössbauer Spectroscopy of Locally Heterogeneous Systems, Almaty, 2000.
57. G. Román-Pérez and J. M. Soler, *Phys. Rev. Lett.* 2009, **103**, 096102.
58. L. Adamska, Y. Lin, A. J. Ross, M. Batzill, and I. I. Oleynik, *Phys. Rev. B* 2012, **85**, 195443.
59. F. Mittendorfer, A. Garhofer, J. redinger, J. Klimeš, J. Harl, and G. Kresse, *Phys. Rev. B* 2011, **84**, 201401(R).
60. H. Zhang, *J. Phys. Chem. Solids* 1999, **60**, 1845.
61. J. Borysiuk, A. Grabias, J. Szczytko, M. Bystrzejewski, A. Twardowski, and H. Lange, *Carbon* 2008, **46**, 1693.
62. T. Enz, M. Winterer, B. Stahl, S. Bhattacharya, G. Miehe, K. Foster, C. Fase, and H. Hahn, *J. Appl. Phys.* 2006, **99**, 044306.
63. S. L. Dudarev, G. A. Botton, S. Y. Savrasov, J. Humphereys, and A. P. Sutton, *Phys. Rev. B* 1998, **57**, 1505.
64. C. Loschen, J. Carrasco, K. M. Neuman, and F. Illas, *Phys. Rev. B* 2007, **75**, 035115.

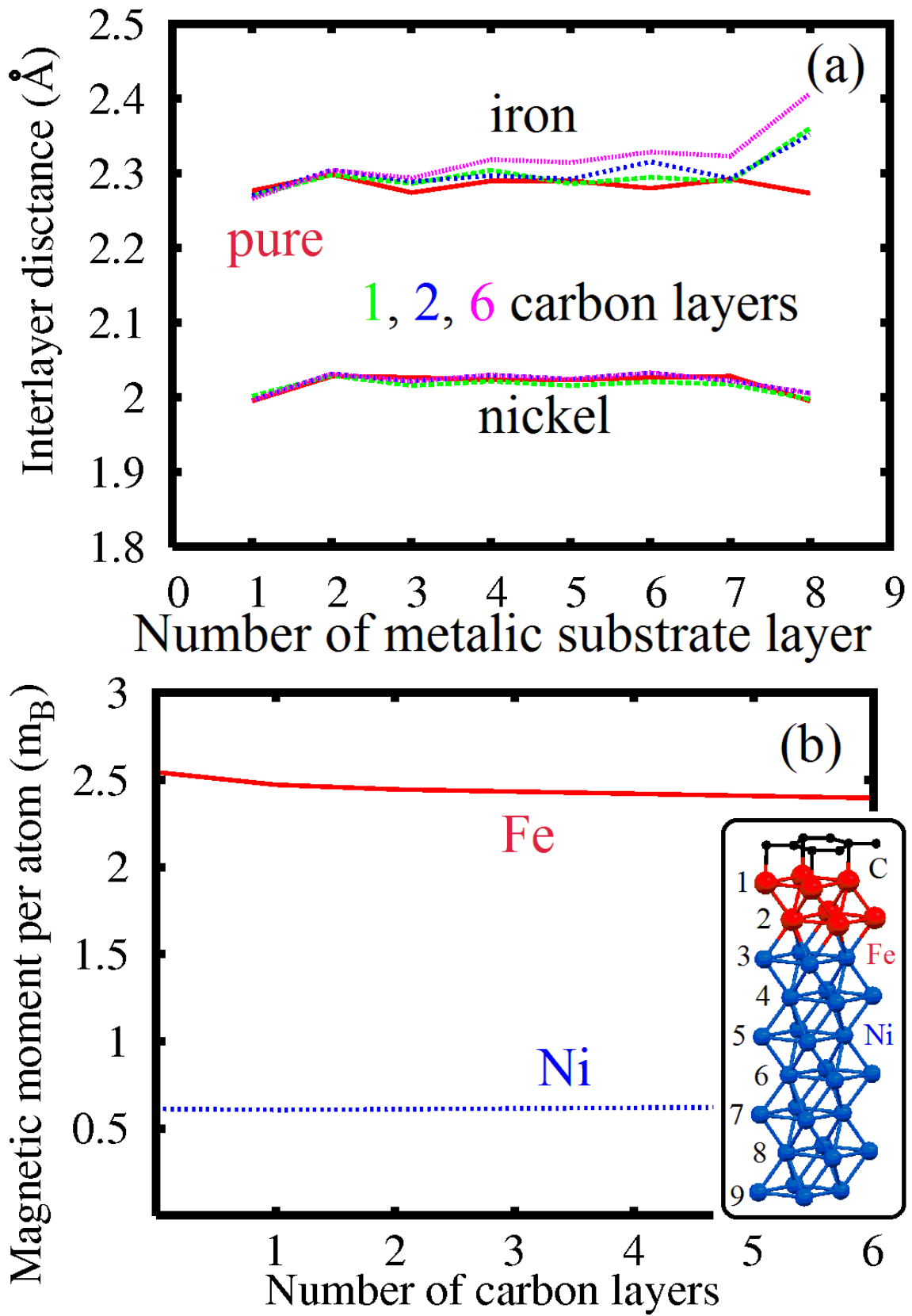

**Figure 1** Interlayer distances in the metallic substrates (a) and average magnetic moment on topmost atoms of metallic substrate interacting with carbon layers (b) as a function of the number of the carbon layers. In the inset is the optimized atomic structure of a graphene monolayer over a nickel substrate capped by two layers of iron.

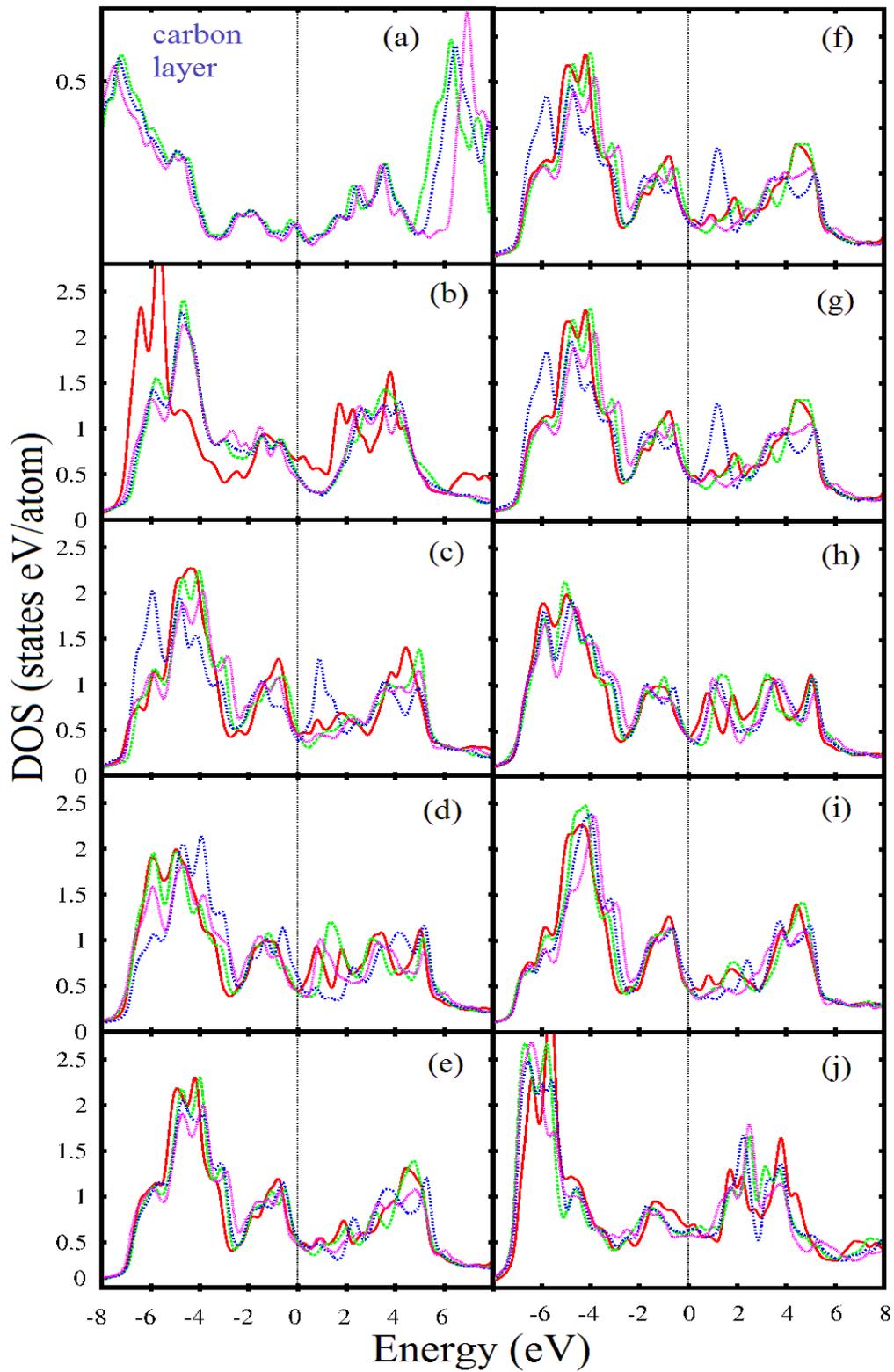

**Figure 2** Local density of states of the carbon atoms over iron substrate (a) and 1$^{st}$-9$^{th}$ (b-j) layers of iron for the cases of the uncoated iron slab (red) and substrate coated with mono- (green), bi- (blue) and six-layer (violet) graphene.

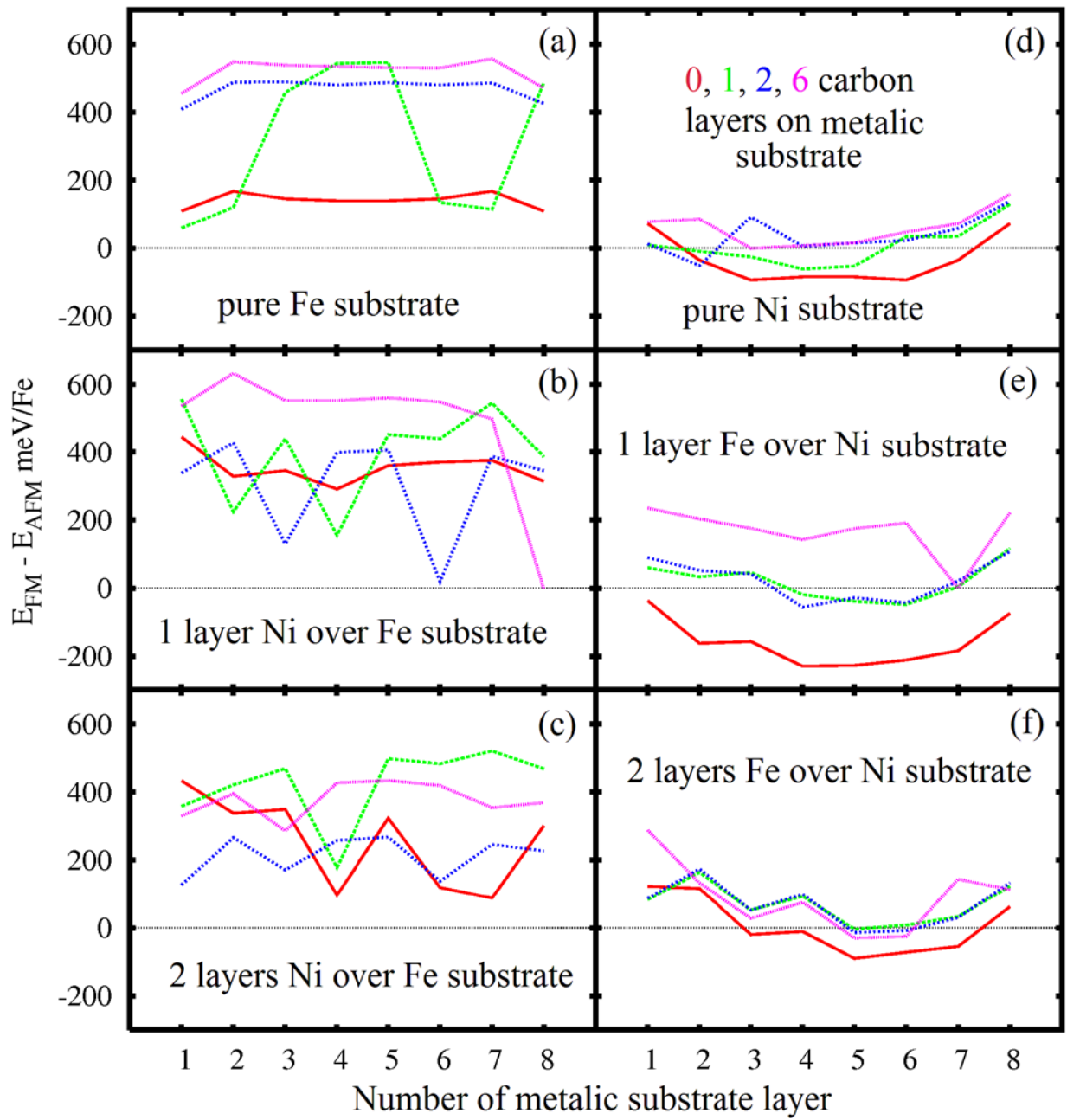

**Figure 3** Values of the exchange interactions between the atoms in various layers of metallic substrates as a function of the substrate composition and the number of graphene layers. The numbering of graphene layers is further: the closest to carbon cover is the first the utmost is 8[th].

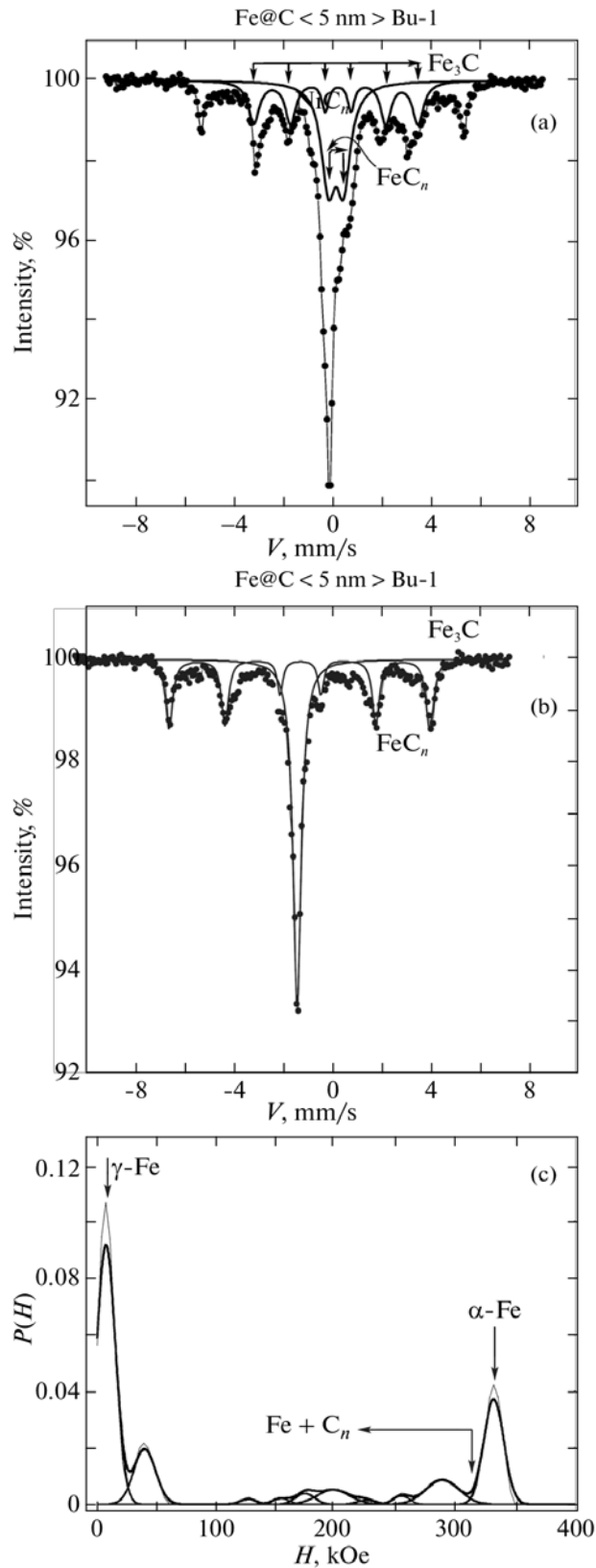

**Figure 4** (a) Mössbauer spectrum of the Fe@C nanocomposite immediately after the synthesis. (b) Spectrum after the subtraction of the phases based on iron carbides and the phase corresponding to the doublet. (c) Function P(H) for this spectrum.

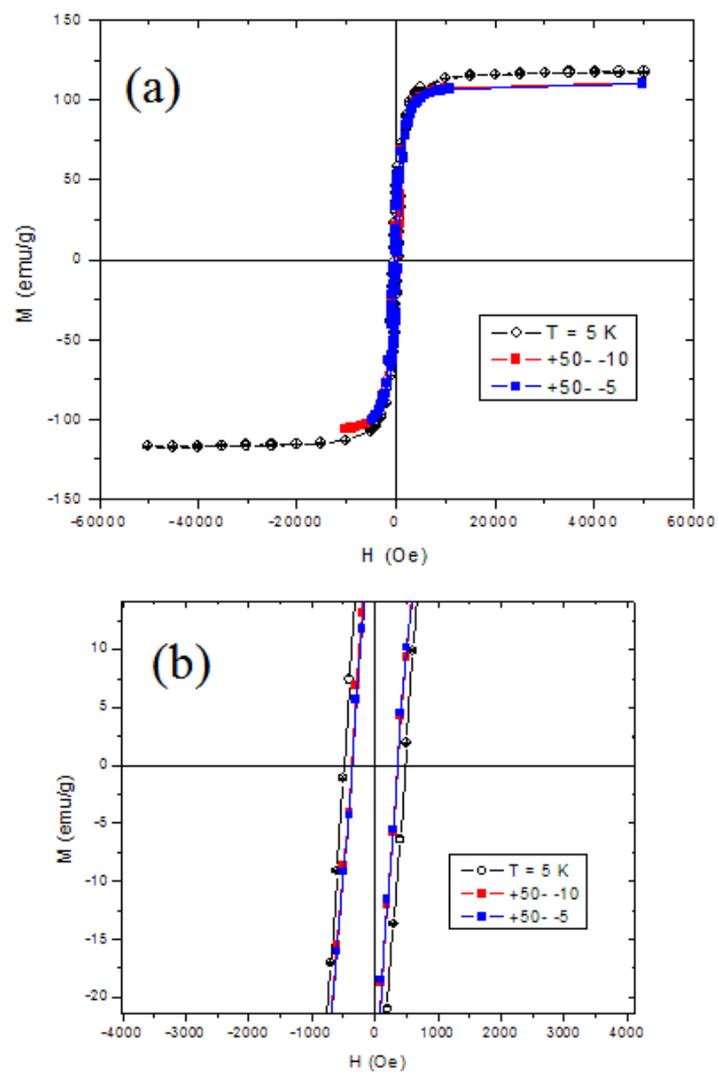

**Figure 5** Magnetization measured at various temperatures for Fe@C nanocomposite samples (synthesized in butane) with an average particle size of 5 nm.

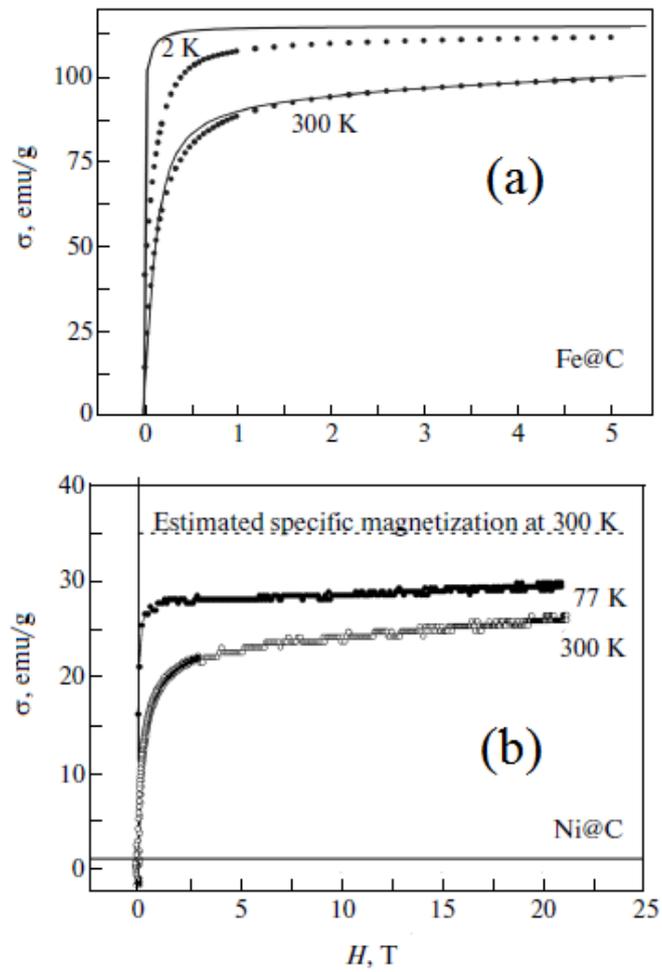

**Figure 6** Magnetization curves of Fe@C (a) and Ni@C (b) nanocomposite samples (synthesized in butane) with average particle sizes of 15 nm. The magnetization was measured at various temperatures in pulsed magnetic fields.

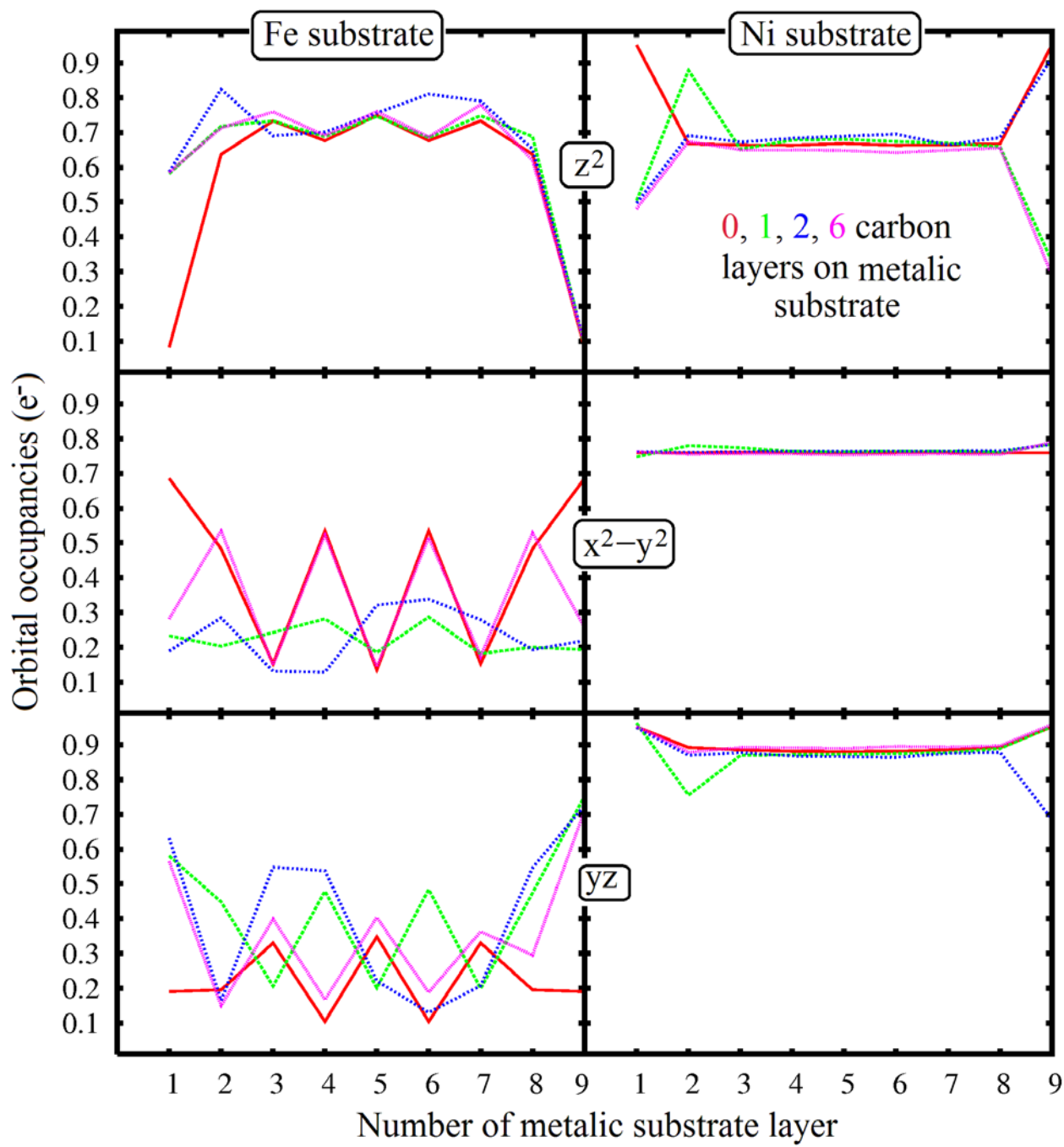

**Figure 7** Occupancies (in electrons) of 'spin-down' (see description in text) 3d orbitals in various layers of uncovered and covered by graphene metallic substrates. The numbering of graphene layers is further: the closest to carbon cover is the first the utmost is 8[th].